**A random laser based on diamond nanoneedles**

Ngoc My Hanh Duong[1], Blake Regan[1], Milos Toth[1], Igor Aharonovich[1] and Judith M Dawes[2]

1. School of Mathematical and Physical Sciences, University of Technology Sydney Ultimo, NSW 2007, Australia

2. MQ Photonics, Department of Physics and Astronomy, Macquarie University, NSW 2019, Australia

E-mail: igor.aharonovich@uts.edu.au; judith.dawes@mq.edu.au

**Abstract**

Random lasers use radiative gain and multiple scatterers in disordered media to generate light amplification. In this study, we demonstrate a random laser based on diamond nanoneedles that act as scatterers in combination with fluorescent dye molecules that serve as a gain medium. Random lasers realized using diamond possess high spectral radiance with angle-free emission and thresholds of 0.16 mJ. The emission dependence on the pillar diameter and density is investigated, and optimum lasing conditions are measured for pillars with spacing and density of ~ 336±40 nm and ~ 2.9 x $10^{10}$ $cm^{-2}$. Our results expand the application space of diamond as a material platform for practical, compact photonic devices and sensing applications.

Diamond has long been considered an ideal material for many applications due to its high chemical stability and thermal conductivity, and excellent optical properties[1] [2] [3] [4] [5]. Indeed diamond has become a robust platform for emerging quantum photonic technologies [6], nonlinear optics[7] [8], optoelectronic devices[9], and biomarking applications[10-12]. In particular, diamond has a high refractive index of 2.4 and optical transparency in the visible part of the spectrum that make it an appealing platform for optical cavities. Indeed, over the last several years, diamond has been used for fabrication of photonic crystal cavities, microdisks, microrings and other optical resonators [13] [14] [15] [16]. In this work, we propose and realize a different concept for diamond photonics by utilizing it as a scattering medium for random lasing.

Random lasing has emerged as a promising candidate for various applications in sensing [17] [18], medical diagnostics [19], and imaging [20] . These lasers are unconventional in that they require no optical cavity but use disordered structures to scatter light. To date, random lasing has been observed in semiconductor powders[21], dielectric scatterers[22, 23], porous silica[24], polymers[25], metallic nanoparticles[26-28], micro and nanowires[29] and optical fibers[30] [31]. Compared to standard lasers which employ optical components such as gratings, waveguides and resonators, the construction and alignment of a random laser is very simple and does not require sophisticated fabrication techniques. Furthermore, the directionality and spatial coherence which are inherent to conventional lasers can be problematic for imaging and display applications. These obstacles motivate research into alternative high efficiency random lasers with simple design and fabrication, low cost and broad angular emission.

To realize a diamond random laser, we engineered disordered nanoscale diamond pillars with a high density using reactive ion etching (RIE), and used a fluorescent dye solution as the gain medium. The diamond nanoneedles act as scatterers for the random lasing process, and also trap the fluorescent dye solution via capillary action and viscous entrainment [32] [33]. These two properties are favourable for the realization of random lasing using an unpatterned mask and fast fabrication process. Figure 1a shows a schematic of our device.

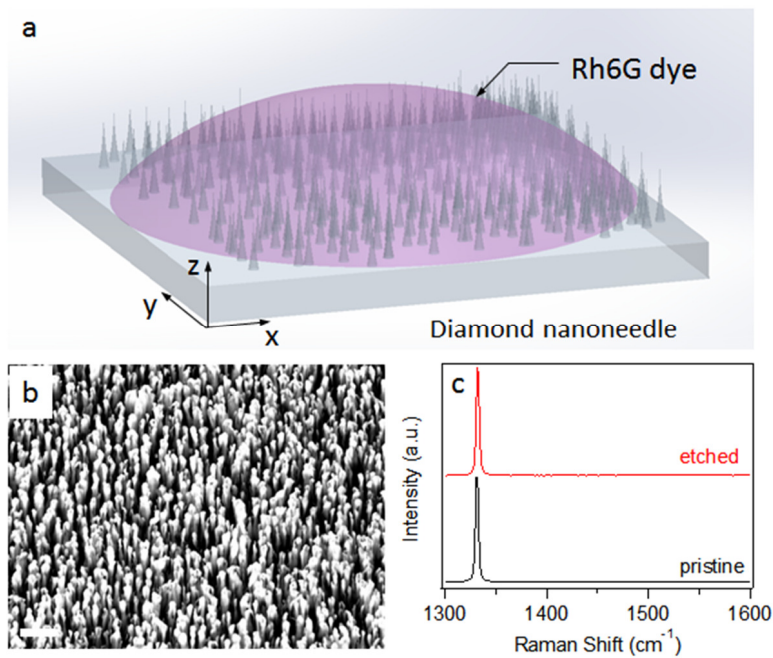

*Figure 1.* *Random laser based on diamond nanoneedles and a fluorescent dye. (a) Schematic illustration of diamond nanoneedles with Rhodamine 6G dye droplet (b) Scanning electron microscope image of sample (i) described in table 1. The scale bar is 1 μm (c) Raman spectra of the diamond before and after etching.*

The main issue with realizing a random laser is the integration of a highly scattering medium with an efficient gain material. We have met these two conditions using diamond nanoneedles infiltrated with a dye, to realise a diamond-based random laser. The generation of pillar arrays by RIE usually requires an etch mask patterned by electron beam lithography or photolithography. However, a random laser does not require an ordered array, and the lithography step can therefore be eliminated. In this study, we spin coated a dilute solution of silica beads containing nanoscale contaminants that act as a disordered etch mask on diamond. Such nano-masking effects result in the formation of highly dense and disordered diamond nanoneedles without the necessity of patterning (figure 1a). The pillar morphology and geometry is seen in figure 1b.

To confirm the pillars were composed of diamond, we performed Raman analysis before and after etching. A laser of 633 nm wavelength and 1800 l/mm gratings was used and only one main Raman active mode at ~ 1332 cm$^{-1}$ was observed. The measured peak was fitted with a Lorentzian function, and the full width at half maximum was <2 cm$^{-1}$. The Raman spectra confirmed that there was no amorphous carbon on the surface of diamond after etching as the presence of carbon would show a peak at 1580 cm$^{-1}$[34].

The emission properties of the diamond nanoneedles containing a gain medium were tested at room temperature by a pulsed laser at normal incidence. A 2 μl droplet of Rhodamine dye 6G 10$^{-3}$ mol/l (Rh6G) was drop cast onto the diamond samples (figure 1a). The dye was dissolved in a methanol: ethylene glycol solution with a ratio of 1:1 to prevent it from drying out during the experiment. The diamond nanoneedles were fabricated by plasma-assisted reactive ion etching, which was followed by a 10 minute clean with O$_2$ plasma to remove carbon byproducts. Usually, submicrometer roughness is known to reduce the liquid-solid contact area and water droplet adhesion, as in the case of the lotus leaf[35-37]. However, a wetting transition occurs on superhydrophobic surfaces at the nanometer scale. The transition between the Cassie−Baxter and Wenzel wetting states occurs as the liquid - solid - air composite interface collapses, resulting in liquid adhering tightly to the surface, and becoming pinned in the structure, which promotes liquid - solid adhesion [38]. The random lasing arises through emission from the dye being scattered by the diamond needles thereby increasing the pathlength in the gain medium and leading to laser action.

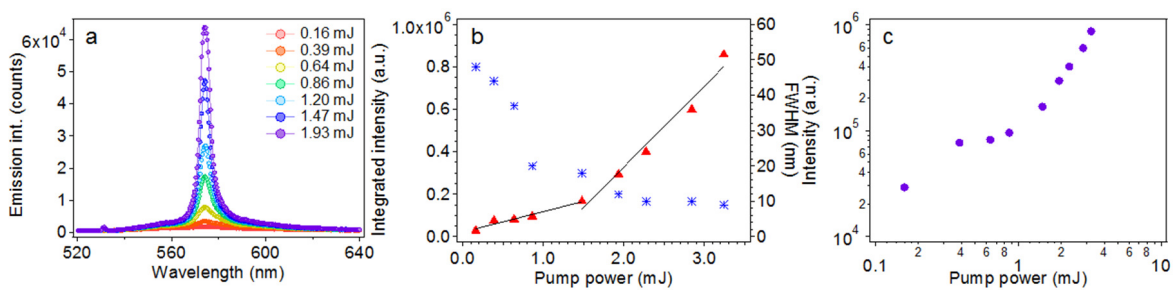

**Figure 2.** *(a) Emission spectra of Rh6G on diamond sample i (table 1) at different pump pulse energies (b) Integrated intensity (red triangles) and linewidth of emission peak (blue asterisks) of Rh6G / diamond nanoneedle random lasers as a function of pump power (c) Log-log plot of integrated intensity at different pump powers showing a characteristic "kink" at the threshold.*

Figure 2a shows the evolution of the emission spectra as the pump energy was increased. At low power, the spectrum showed broad emission, with no random lasing. As the pump energy was increased, the emission peak became narrower at ~ 573 nm due to preferential amplification at frequencies close to the maximum of the gain spectrum, as can be seen from figure 2b. The linewidth of the lasing peak was ~ 9 nm, which was 6 times narrower than the linewidth of the amplified spontaneous emission peak below the threshold. A threshold behavior was observed at ~ 0.64 mJ as the integrated intensity increases nonlinearly with the pump power[39, 40]. The lasing threshold is clearly confirmed as a "kink" in the log-log plot (figure 2c). The nature of lasing behaviour in a disordered gain medium is the result of recurrent scattering events happening in strong scattering and gain media. Because of the short scattering mean free path in the diamond nanoneedles, the light is strongly scattered and closed loop paths can be created through multiple scattering, yielding laser oscillation. When the pump intensity increases, the gain of the optical modes increases to be equal to cavity loss. When the optical gain exceeds the loss, laser oscillation occurs in these cavities.

The photon mean free path is an important parameter in random laser systems and is determined by the nature and the density of the scatterers. By changing the etching parameters, we have the flexibility to vary the density of the nanoneedles. Table 1 summarises the etching conditions that have been used to treat the diamond samples.

| Sample | ICP (W) | RIE (W) | O$_2$/Ar (sccm) | Spacing (nm) | Width (nm) | Taper angle |
|--------|---------|---------|-----------------|--------------|------------|-------------|
| i | 500 | 100 | 45/5 | 325±23 | 86±8 | 81±2 |
| ii | 500 | 100 | 45/20 | 282±21 | 65±4 | 90±2 |
| iii | 300 | 100 | 45/5 | 177±11 | 88±7 | 89±2 |
| iv | 500 | 50 | 45/5 | 336±38 | 139±15 | 67±4 |

**Table 1.** *Specific reactive ion etching conditions used on different diamond samples. The average values for the spacing, width and taper angle were measured from the SEM images in figure 1b and figure 3a, b, c.*

The etch conditions (Table 1) were designed to follow a standard diamond etch procedure, with variations[41, 42] controlling the rate of non-chemical etching to control inorganic contamination responsible for the masking of the diamond pillars. By increasing the nonreactive ratio of the etch gas (Condition ii), the physical etch is more capable of etching minor surface contamination and re-sputtering the contamination across the surface more evenly, resulting in greater needle density (figure 3a). By decreasing the ICP bias (condition iii) we lower the concentration of active ions, resulting in a higher physical etch and re-sputtering rate, causing a higher density of nanoneedles. This condition also results in fewer reactive species that yield an anisotropic etching profile, resulting in a straighter needle morphology as seen in figure 3b. The decreased RIE bias (Condition iv) causes chemical etching to dominate, resulting in minimal wear of surface masking and thus a wider distribution of the nanoneedles as seen in figure 3c. The lowered RIE bias also increases the rate of isotropic etching induced on the side walls, forming a more tapered structure with sharp sidewalls.

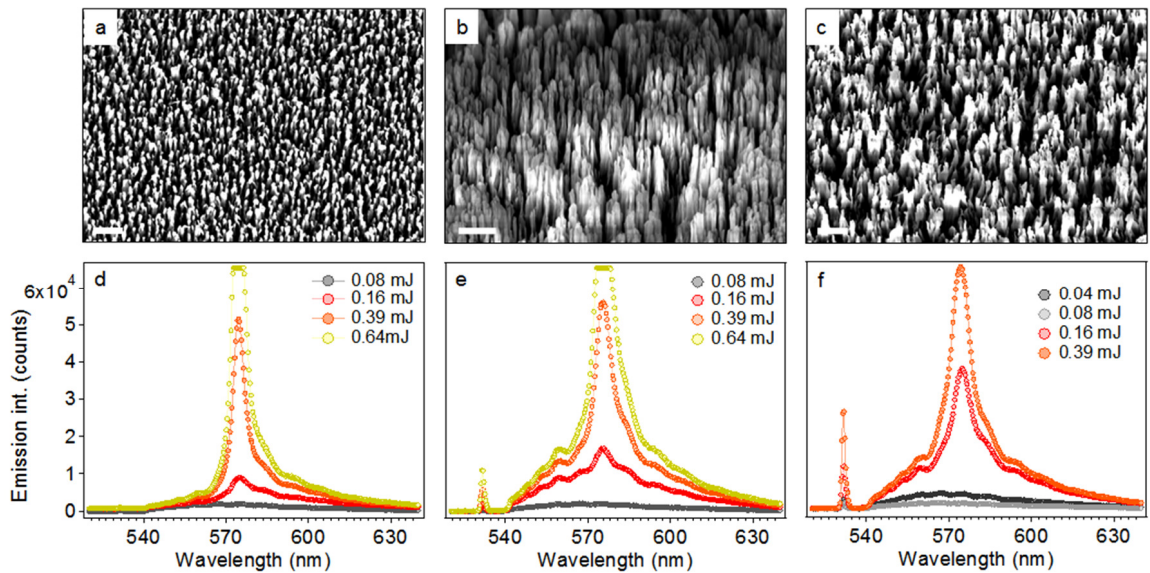

**Figure 3.** *Different morphologies of random diamond nanoneedles from samples with varying etching conditions and emission spectra. SEM images of diamond samples etched using different conditions summarised in Table 1 (a) Sample ii (b) Sample iii and (c) Sample iv. The scale bar is 1 um. Emission spectra of Rh6G dye solution 10⁻³ mol/l when the pump energy is 0.08, 0.16, 0.39 and 0.64 mJ, respectively, on different diamond samples.*

A larger needle spacing results in longer scattering mean free paths, which helps to yield the lowest lasing threshold in sample c ~ 0.16 mJ (figure 4c) compared to the higher thresholds in samples a and b, which are ~ 0.39 mJ. Because of the high density of nanoneedles, the photon paths are typically shorter. Hence, the threshold pumping energy increased (figure 4 a, b) [39]. Another contribution to the decrease in the threshold is the nanoneedle geometry. Nanopillars with tapered profiles can suppress the reflectance of the flat substrate by gradually changing the refractive index contrast, thus incident light would concentrate in the disordered substrate. Enhanced absorption would then increase the optical gain efficiency of dye and correspondingly amplify the spontaneous emission[43].

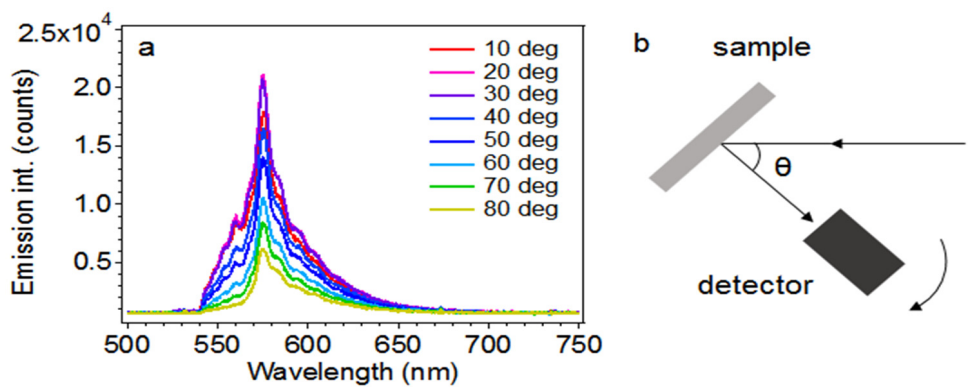

**Figure 4.** *(a) Emission spectra of Rhodamine 6G solution on diamond nanoneedles collected at different detector angles from the sample surface. The pump energy was 0.16 mJ. (b) Schematic of the measurement. θ is the rotation angle between the incident pump and emission light.*

Finally, random laser emission from the Rh6G/nanoneedles should be observed in all directions. Figure 4a shows the lasing emission spectra observed at different collection angles. Since different laser cavities formed by multiple scattering events could have different output directions, lasing modes observed at different angles can, in principle, be different. In our case, due to a 5 ns pulse width of the excitation laser, the laser peaks were averaged and merged into one broad peak.

We successfully realized random lasers from diamond nanoneedles integrated with Rhodamine dye solution. We found that the best diamond nanoneedle sample for the random lasers is sample iv, containing tapered and sharp edge needles with average width of 139±15 nm, exhibits the lowest threshold of 0.16 mJ. Future directions of this work may include integration of diamond nanoneedles with microfluidic channels to achieve tuneable lasing and testing of other gain media. Our results are promising for applications in imaging, sensing and display technologies.

## Experimental Section

*Sample preparation:*

The single crystal diamond sample (3 mm x 3 mm x 0.3 mm, Element Six) had the top face (100) synthesized by chemical vapour deposition (CVD) with nitrogen concentration < 1 ppm. A dilute silica bead solution was ultra-sonicated and vortex-mixed to evenly distribute the silicon nanoparticles in solution. A 0.5 mL droplet was dropcast on the diamond sample. The diamond samples were flushed with $O_2$ plasma for 5 minutes before the RIE process using inductively coupled plasma reactive ion etching (ICP-RIE), followed by a 10 minute $O_2$ plasma clean.

*Optical pumping experiments:*

A Q-switched, frequency-doubled Nd:YAG laser (532 nm wavelength, 1 to 10 Hz repetition rate, 5 ns pulse width) with a 3 mm pump beam diameter was used to pump the samples at an angle of 45° to the normal to the front face of the sample, which was adhered to a glass slide by immersion oil and the slide was mounted vertically. The dye liquid remained in position during the experiments, due to its viscosity and the entrainment by the nanoneedles [32] [33]. A lens (f = 5 cm) was placed at 30° to the normal in front of the sample to collect emission light to a fibre-coupled spectrometer (Ocean Optics, USB2000 UV-VS-ES ~1 nm resolution). A 532 nm edge filter was used to block residual pump light from the spectrometer.

## Acknowledgements

Financial support from the Australian Research council (via DP140102721, DP180100077, LP170100150), the Asian Office of Aerospace Research and Development grant FA2386-17-1-4064, the Office of Naval Research Global under grant number N62909-18-1-2025 are gratefully acknowledged.

## References

[1] J.W. Liu, M.Y. Liao, M. Imura, R.G. Banal, Y. Koide, Deposition of TiO2/Al2O3 bilayer on hydrogenated diamond for electronic devices: Capacitors, field-effect transistors, and logic inverters, Journal of Applied Physics, 121 (2017) 224502.

[2] K.G. Crawford, D. Qi, J. McGlynn, T.G. Ivanov, P.B. Shah, J. Weil, A. Tallaire, A.Y. Ganin, D.A.J. Moran, Thermally Stable, High Performance Transfer Doping of Diamond using Transition Metal Oxides, Scientific Reports, 8 (2018) 3342.

[3] J.L. Zhang, S. Sun, M.J. Burek, C. Dory, Y.-K. Tzeng, K.A. Fischer, Y. Kelaita, K.G. Lagoudakis, M. Radulaski, Z.-X. Shen, N.A. Melosh, S. Chu, M. Lončar, J. Vučković, Strongly Cavity-Enhanced Spontaneous Emission from Silicon-Vacancy Centers in Diamond, Nano Letters, 18 (2018) 1360-1365.

[4] L. Yates, J. Anderson, X. Gu, C. Lee, T. Bai, M. Mecklenburg, T. Aoki, M.S. Goorsky, M. Kuball, E.L. Piner, S. Graham, Low Thermal Boundary Resistance Interfaces for GaN-on-Diamond Devices, ACS Applied Materials & Interfaces, 10 (2018) 24302-24309.

[5] D.P. Lake, M. Mitchell, Y. Kamaliddin, P.E. Barclay, Optomechanically Induced Transparency and Cooling in Thermally Stable Diamond Microcavities, ACS Photonics, 5 (2018) 782-787.

[6] I. Aharonovich, A.D. Greentree, S. Prawer, Diamond photonics, Nature Photonics, 5 (2011) 397.

[7] B.J.M. Hausmann, I. Bulu, V. Venkataraman, P. Deotare, M. Lončar, Diamond nonlinear photonics, Nature Photonics, 8 (2014) 369.

[8] R.P. Mildren, J.E. Butler, J.R. Rabeau, CVD-diamond external cavity Raman laser at 573 nm, Opt. Express, 16 (2008) 18950-18955.

[9] L. Li, Y. Yu, G.J. Ye, Q. Ge, X. Ou, H. Wu, D. Feng, X.H. Chen, Y. Zhang, Black phosphorus field-effect transistors, Nat Nanotechnol, 9 (2014) 372-377.

[10] N. Mohan, Y.-K. Tzeng, L. Yang, Y.-Y. Chen, Y.Y. Hui, C.-Y. Fang, H.-C. Chang, Sub-20-nm Fluorescent Nanodiamonds as Photostable Biolabels and Fluorescence Resonance Energy Transfer Donors, Advanced Materials, 22 (2010) 843-847.

[11] Y.-R. Chang, H.-Y. Lee, K. Chen, C.-C. Chang, D.-S. Tsai, C.-C. Fu, T.-S. Lim, Y.-K. Tzeng, C.-Y. Fang, C.-C. Han, H.-C. Chang, W. Fann, Mass production and dynamic imaging of fluorescent nanodiamonds, Nature Nanotechnology, 3 (2008) 284.

[12] N. Mohan, C.-S. Chen, H.-H. Hsieh, Y.-C. Wu, H.-C. Chang, In Vivo Imaging and Toxicity Assessments of Fluorescent Nanodiamonds in Caenorhabditis elegans, Nano Letters, 10 (2010) 3692-3699.

[13] T.M. Babinec, B.J.M. Hausmann, M. Khan, Y. Zhang, J.R. Maze, P.R. Hemmer, M. Lončar, A diamond nanowire single-photon source, Nature Nanotechnology, 5 (2010) 195.

[14] J. Riedrich-Möller, L. Kipfstuhl, C. Hepp, E. Neu, C. Pauly, F. Mücklich, A. Baur, M. Wandt, S. Wolff, M. Fischer, S. Gsell, M. Schreck, C. Becher, One- and two-dimensional photonic crystal microcavities in single crystal diamond, Nature Nanotechnology, 7 (2011) 69.

[15] J.C. Lee, I. Aharonovich, A.P. Magyar, F. Rol, E.L. Hu, Coupling of silicon-vacancy centers to a single crystal diamond cavity, Opt. Express, 20 (2012) 8891-8897.

[16] I. Aharonovich, J.C. Lee, A.P. Magyar, B.B. Buckley, C.G. Yale, D.D. Awschalom, E.L. Hu, Homoepitaxial Growth of Single Crystal Diamond Membranes for Quantum Information Processing, Advanced Materials, 24 (2012) OP54-OP59.

[17] D.S. Wiersma, S. Cavalieri, A temperature-tunable random laser, Nature, 414 (2001) 708.

[18] W.Z. Wan Ismail, G. Liu, K. Zhang, E.M. Goldys, J.M. Dawes, Dopamine sensing and measurement using threshold and spectral measurements in random lasers, Opt. Express, 24 (2016) A85-A91.

[19] R. Choe, A. Corlu, K. Lee, T. Durduran, S.D. Konecky, M. Grosicka-Koptyra, S.R. Arridge, B.J. Czerniecki, D.L. Fraker, A. DeMichele, B. Chance, M.A. Rosen, A.G. Yodh, Diffuse optical tomography of breast cancer during neoadjuvant chemotherapy: A case study with comparison to MRI, Medical Physics, 32 (2005) 1128-1139.

[20] B. Redding, M.A. Choma, H. Cao, Speckle-free laser imaging using random laser illumination, Nature Photonics, 6 (2012) 355.

[21] H. Cao, Y.G. Zhao, X. Liu, E.W. Seelig, R.P.H. Chang, Effect of external feedback on lasing in random media, Applied Physics Letters, 75 (1999) 1213-1215.

[22] H. Cao, Y.G. Zhao, S.T. Ho, E.W. Seelig, Q.H. Wang, R.P.H. Chang, Random Laser Action in Semiconductor Powder, Physical Review Letters, 82 (1999) 2278-2281.

[23] H. Fujiwara, R. Niyuki, Y. Ishikawa, N. Koshizaki, T. Tsuji, K. Sasaki, Low-threshold and quasi-single-mode random laser within a submicrometer-sized ZnO spherical particle film, Applied Physics Letters, 102 (2013) 061110.

[24] M.V. Santos, É. Pecoraro, S.H. Santagneli, A.L. Moura, M. Cavicchioli, V. Jerez, L.A. Rocha, L.F.C. de Oliveira, A.S.L. Gomes, C.B. de Araújo, S.J.L. Ribeiro, Silk fibroin as a biotemplate for hierarchical porous silica monoliths for random laser applications, Journal of Materials Chemistry C, 6 (2018) 2712-2723.

[25] S. Li, L. Wang, T. Zhai, J. Tong, L. Niu, F. Tong, F. Cao, H. Liu, X. Zhang, A dual-wavelength polymer random laser with the step-type cavity, Organic Electronics, 57 (2018) 323-326.

[26] Y.-C. Yao, Z.-P. Yang, J.-M. Hwang, H.-C. Su, J.-Y. Haung, T.-N. Lin, J.-L. Shen, M.-H. Lee, M.-T. Tsai, Y.-J. Lee, Coherent and Polarized Random Laser Emissions from Colloidal CdSe/ZnS Quantum Dots Plasmonically Coupled to Ellipsoidal Ag Nanoparticles, Advanced Optical Materials, 5 (2016) 1600746.

[27] T. Zhai, Z. Xu, X. Wu, Y. Wang, F. Liu, X. Zhang, Ultra-thin plasmonic random lasers, 2016.

[28] T. Zhai, X. Zhang, Z. Pang, X. Su, H. Liu, S. Feng, L. Wang, Random Laser Based on Waveguided Plasmonic Gain Channels, Nano Letters, 11 (2011) 4295-4298.

[29] Y. Ren, H. Zhu, Y. Wu, G. Lou, Y. Liang, S. Li, S. Su, X. Gui, Z. Qiu, Z. Tang, Ultraviolet Random Laser Based on a Single GaN Microwire, ACS Photonics, 5 (2018) 2503-2508.

[30] J. He, W.-K. Chan, X. Cheng, M.-L. Tse, C. Lu, P.-K. Wai, S. Savovic, H.-Y. Tam, Experimental and Theoretical Investigation of the Polymer Optical Fiber Random Laser with Resonant Feedback, Advanced Optical Materials, 6 (2018) 1701187.

[31] F. Luan, B. Gu, A.S.L. Gomes, K.-T. Yong, S. Wen, P.N. Prasad, Lasing in nanocomposite random media, Nano Today, 10 (2015) 168-192.

[32] A. Nasto, P.T. Brun, A.E. Hosoi, Viscous entrainment on hairy surfaces, Physical Review Fluids, 3 (2018) 024002.

[33] C. Ishino, M. Reyssat, E. Reyssat, K. Okumura, D. Quéré, Wicking within forests of micropillars, EPL (Europhysics Letters), 79 (2007) 56005.

[34] T. Lühmann, R. Wunderlich, R. Schmidt-Grund, J. Barzola-Quiquia, P. Esquinazi, M. Grundmann, J. Meijer, Investigation of the graphitization process of ion-beam irradiated diamond using ellipsometry, Raman spectroscopy and electrical transport measurements, Carbon, 121 (2017) 512-517.

[35] K. Koch, W. Barthlott, Superhydrophobic and superhydrophilic plant surfaces: an inspiration for biomimetic materials, Philosophical Transactions of the Royal Society A: Mathematical, Physical and Engineering Sciences, 367 (2009) 1487-1509.

[36] B. Bhushan, Y.C. Jung, K. Koch, Micro-, nano- and hierarchical structures for superhydrophobicity, self-cleaning and low adhesion, Philosophical Transactions of the Royal Society A: Mathematical, Physical and Engineering Sciences, 367 (2009) 1631-1672.

[37] B. Bhushan, Y.C. Jung, K. Koch, Self-Cleaning Efficiency of Artificial Superhydrophobic Surfaces, Langmuir, 25 (2009) 3240-3248.

[38] H. Teisala, M. Tuominen, M. Aromaa, M. Stepien, J.M. Mäkelä, J.J. Saarinen, M. Toivakka, J. Kuusipalo, Nanostructures Increase Water Droplet Adhesion on Hierarchically Rough Superhydrophobic Surfaces, Langmuir, 28 (2012) 3138-3145.

[39] J. Kitur, G. Zhu, M. Bahoura, M.A. Noginov, Dependence of the random laser behavior on the concentrations of dye and scatterers, Journal of Optics, 12 (2010) 024009.

[40] D.S. Wiersma, The physics and applications of random lasers, Nature Physics, 4 (2008) 359.

[41] B.J.M. Hausmann, M. Khan, Y. Zhang, T.M. Babinec, K. Martinick, M. McCutcheon, P.R. Hemmer, M. Lončar, Fabrication of diamond nanowires for quantum information processing applications, Diamond and Related Materials, 19 (2010) 621-629.

[42] B. Khanaliloo, M. Mitchell, A.C. Hryciw, P.E. Barclay, High-Q/V Monolithic Diamond Microdisks Fabricated with Quasi-isotropic Etching, Nano Letters, 15 (2015) 5131-5136.

[43] R. Sanatinia, K.M. Awan, S. Naureen, N. Anttu, E. Ebraert, S. Anand, GaAs nanopillar arrays with suppressed broadband reflectance and high optical quality for photovoltaic applications, Opt. Mater. Express, 2 (2012) 1671-1679.